\documentclass[dvipdfmx,11pt,a4]{article}
\usepackage[top=20truemm,bottom=20truemm,left=20truemm,right=20truemm]{geometry}
\usepackage[dvipdfmx]{graphicx,color}
\usepackage{amssymb,amsmath}
\usepackage{hyperref}
\hypersetup{
setpagesize=false,
 bookmarksnumbered=true,%
 bookmarksopen=true,%
 colorlinks=true,%
 linkcolor=black,
 citecolor=black,
}
\newcommand{\eq}[1]{\mbox{Eq.~(\ref{#1})}}
\newcommand{\eqs}[1]{\mbox{Eqs.~(\ref{#1})}}
\newcommand{\fig}[1]{\mbox{Fig.~\ref{#1}}}
\newcommand{\D}{{\rm d}}
\newcommand\ringring[1]{%
  {
   \mathop{\kern0pt #1}\limits^{
     \vbox to-1.85ex{
       \kern-2ex 
       \hbox to 0pt{\hss\normalfont\kern.1em \r{}\kern-.45em \r{}\hss}%
       \vss 
     }
   }
  }
}

\begin{document}

\vfill
\begin{flushright}
KEK-Cosmo-218,
KEK-TH-2022,
RUP-17-25
\end{flushright}

\begin{center}
{\Large{\bf Energy emission from high curvature region and its backreation
}}

\vskip 1cm
{\bf
Takafumi Kokubu$^{1,4,*}$, Sanjay Jhingan$^{2, 3, \dagger}$ and Tomohiro Harada$^{4,\S}$\\
}

\vskip 1cm
{ 
{\it $^1$Theory Center, Institute of Particle and Nuclear Studies, High Energy Accelerator Research Organization (KEK), Tsukuba 305-0801, Japan}\\
{\it $^2$ iCLA, Yamanashi Gakuin University, Kofu, Yamanashi, 400-8575, Japan}\\
{\it $^3$ Centre for Theoretical Physics, Jamia Millia Islamia, New Delhi 110025, India}\\
{\it $^4$ Department of Physics, Rikkyo University, Toshima, Tokyo, 171-8501, Japan}\\
$^*$kokubu@post.kek.jp,  14ra002a@rikkyo.ac.jp, $^\dagger$ sanjay.jhingan@gmail.com, $^\S$harada@rikkyo.ac.jp
}

{\bf Abstract}
 \end{center}

A strong gravity naked singular region can give important clues towards understanding classical as well as spontaneous nature of General Relativity. We propose here a model for energy emission from a naked singular region in a self-similar dust spacetime by gluing two self-similar dust solutions at the Cauchy horizon. The energy is defined and evaluated as a surface energy of a null hypersurface, the null shell. Also included are scenarios of spontaneous creation or disappearance of a singularity, end of inflation, black hole formation and bubble nucleation. 
Our examples investigated here explicitly show that one can model unlimitedly luminous and energetic objects in the framework of General Relativity.

\section{Introduction}

The recent discoveries of gravitational waves \cite{GW150914, GW151226, GW170104, GW170814, GW170817} has provided us with opportunity to study strong gravitational field environments created by coalescing binary black holes and neutron stars. Having established General Relativity (GR) as the classical theory in strong gravity regime these discoveries has also opened new exciting areas research such as gravitational wave astronomy and multi-messenger astronomy \cite{mmess}. However, theoretically, GR predicts that under quite generic initial conditions continued gravitational collapse should end in a singularity, i.e. regions of arbitrarily high spacetime curvature \cite{pen65, Haw67, reviews}. Thus black holes as unique endstate of continued gravitational collapse is contingent on an additional requirement of cosmic censorship conjecture (CCC) \cite{penrose1969}. The essence of this conjecture is that the singularity should always be safely hidden behind a horizon. This can be achieved by formation of trapped surfaces which form during collapse and surround the singularity hiding it from the outside world. One could, perhaps, also argue that before mathematical structure of spacetime breaks down some quantum gravity effects, due to breakdown of classical theory, should prevent singularity formation. However, with no viable quantum theory for gravity in sight the physics of such dense regions in spacetime remains an open question. Interestingly GR all by itself does not say anything about causal nature of these singularities. Therefore, agreement between GR and strong gravity observations has also renewed interest in the formation of naked singularities as an additional possibility of the end state of gravitational collapse.

The study by Oppenheimer and Snyder (OS) provides a paradigm for black hole formation in continued gravitational collapse \cite{Oppenheimer-Snyder}. This simple model of a spherically symmetric homogeneous dust ball (perfect fluid with no pressure) elegantly captures salient features characterizing a black hole, namely, the formation of a central singularity, an apparent and an event horizon. The OS model led to the establishment viewpoint that end-state of continued gravitational collapse leads to the formation of a black hole (CCC). Numerous counter-examples to CCC have also appeared in the studies generalizing the OS model, i.e., formation of a singularity which can be seen by an external observer. Such a naked singularity can form as an end state of collapsing inhomogeneous dust cloud \cite{eardley-smarr1978, roberts1989, Ori-piran1987, Christodoulou1984, Joshi-Dwivedi1993}. There are radial null geodesics starting from the singularity. With an appropriate choice of initial data, null geodesics can reach the null infinity, thus the singularity can be globally naked.

The counterexamples to censorship hypothesis are not limited to dust only. Gravitational collapse in matter models with pressure are investigated as well. The spherically symmetric self-similar spacetime filled with a perfect fluid forms a  naked singularity if an equation of state is soft enough \cite{Ori-piran1987}. Furthermore, naked singularity also forms when assumption of self-similarity is relaxed\cite{Harada98, HaradaMaeda2001} (see also \cite{coop}).
The generalization of geometry to quasi-spherical or cylindrical also does not help restoring CCC \cite{Qs-CS}. 
Thus, the studies so far indicate that arbitrarily high curvature regions, akin to a naked singularity, can form during the collapse.

The extreme curvature region around the singularity is expected to give rise to high energy phenomena. Since there is virtually no upper limit to energy that can be achieved due to strong curvatures, high energetic phenomena in the universe, such as gamma-ray burst, can be from a region around a strong curvature naked singularity. 
However, these studies are speculative in nature and it is a challenge to create a concrete model that can serve to explain the origin of high energy phenomena like a gamma-ray burst. 
There are few qualitative \cite{ChakrabartiJoshi1994, Singh1998} and quantitative studies \cite{HiscockWilliamsEardley1982} - \cite{HaradaIguchiNakao2002}  in this direction.  Classical gravitational radiation of spacetimes filled with dust matter was studied in \cite{IguchiNakaoHarada1998, IguchiHaradaNakao1999, IguchiHaradaNakao2000}.

As mentioned above, a self-similar dust universe with an appropriate initial matter distribution forms a naked singularity. The fastest of these geodesics from the singular center defines the Cauchy horizon. In a spherically symmetric spacetime geometry, starting from the singular center, the Cauchy horizon expands radially and the region beyond the horizon is by definition undetermined. Cauchy horizon being a null hypersurface we use the null version of junction conditions to replace the inside of Cauchy horizon by another well behaved spacetime removing singularity and restoring predictability.
 
The aim of this paper is to propose various models for physical processes such as energy emission from singular regions, spontaneous creation of a singularity, negative mass black holes, decay of Minkowski spacetime, a scenario for end of inflation, and spontaneous creation of black holes, using matching across two null hypersurfaces. The matching across two null hypersurfaces was pioneered by W. Israel and collaborators (see \cite{Barrabes-Hogan-book}, and references therein). To achieve our goal we use dust models which are known to harbor nakedly singular solutions.   


The paper is organized as follows. In Sec.\ref{General formalism for spherically symmetric null shell}, we give a brief overview of null-shell formalism describing an expanding or contracting null surface which partitions a spacetime into two parts. Sec. \ref{Examples} contains several examples based on the general formalism. We conclude with Sec.\ref{Conclusion}. 
Units $c=G=1$ are adopted throughout this paper.

\section{Spherically symmetric null shell - general formalism}\label{General formalism for spherically symmetric null shell}
We give a quick overview of general formalism that describes an expanding or imploding null hypersurface with a finite surface energy and pressure in
spherically symmetric spacetimes. Whereas the spherically symmetric case is discussed in \cite{toolkit}, the analysis and notation used here are in view of the examples discussed later in the paper. The formalism models the energy and the pressure as the quantities defined on the boundary of two matched spacetimes. The goal here is to calculate the energy that such a null surface ($\Sigma$) would have.   
The spacetime metrics on either side of $\Sigma$ are given by
\begin{align}
\D s_\pm^2=A_\pm(t_\pm,r_\pm)\D t_\pm^2+B_\pm(t_\pm,r_\pm)\D r_\pm^2+C_\pm(t_\pm,r_\pm)(\D \theta^2+\sin^2\D \varphi^2) . \label{general-metric}
\end{align}
Here, $4\pi C(t,r)$ is equivalent to physical surface area of a sphere
at coordinate $t$ and $r$.  We identify spacetimes on the either side
of null hypersurface ($\Sigma$) by $+$ and $-$  signs. 

For completeness we give here the non-zero Christoffel symbols for the metric above ($\pm$ sign is omitted):
\begin{align}
& \Gamma^t_{tt}=\frac{\dot A}{2A},~ \Gamma^t_{tr}=\frac{A'}{2A},~ \Gamma^t_{rr}=-\frac{\dot B}{2A},~ \Gamma^t_{\theta \theta}=-\frac{\dot C}{2A},~ \Gamma^t_{\varphi \varphi}=-\frac{\dot C}{2A}\sin^2\theta, \nonumber \\
& \Gamma^r_{rr}=\frac{B'}{2B},~\Gamma^r_{tr}=\frac{\dot B}{2B},~\Gamma^r_{tt}=-\frac{A'}{2B},~\Gamma^r_{\theta \theta}=-\frac{C'}{2B},~\Gamma^r_{\varphi \varphi}=-\frac{C'}{2B}\sin^2\theta, \nonumber \\
& \Gamma^\theta_{t \theta}=\frac{\dot C}{2C},~\Gamma^\theta_{r \theta}=\frac{C'}{2C},~\Gamma^\theta_{\varphi \varphi}=-\sin \theta \cos \theta,~
 \Gamma^\varphi_{t \varphi}=\frac{\dot C}{2C},~\Gamma^\varphi_{r\varphi}=\frac{C'}{2C},~\Gamma^\varphi_{\theta \varphi}=\cot \theta, \label{christoffel-general-sph}
\end{align}
where $\cdot:=\partial/\partial t$ and $':=\partial/\partial r$.

\subsection{The null hypersurface $\Sigma$}
\begin{figure}[htb]
\begin{center}
\includegraphics[clip,width=10cm]{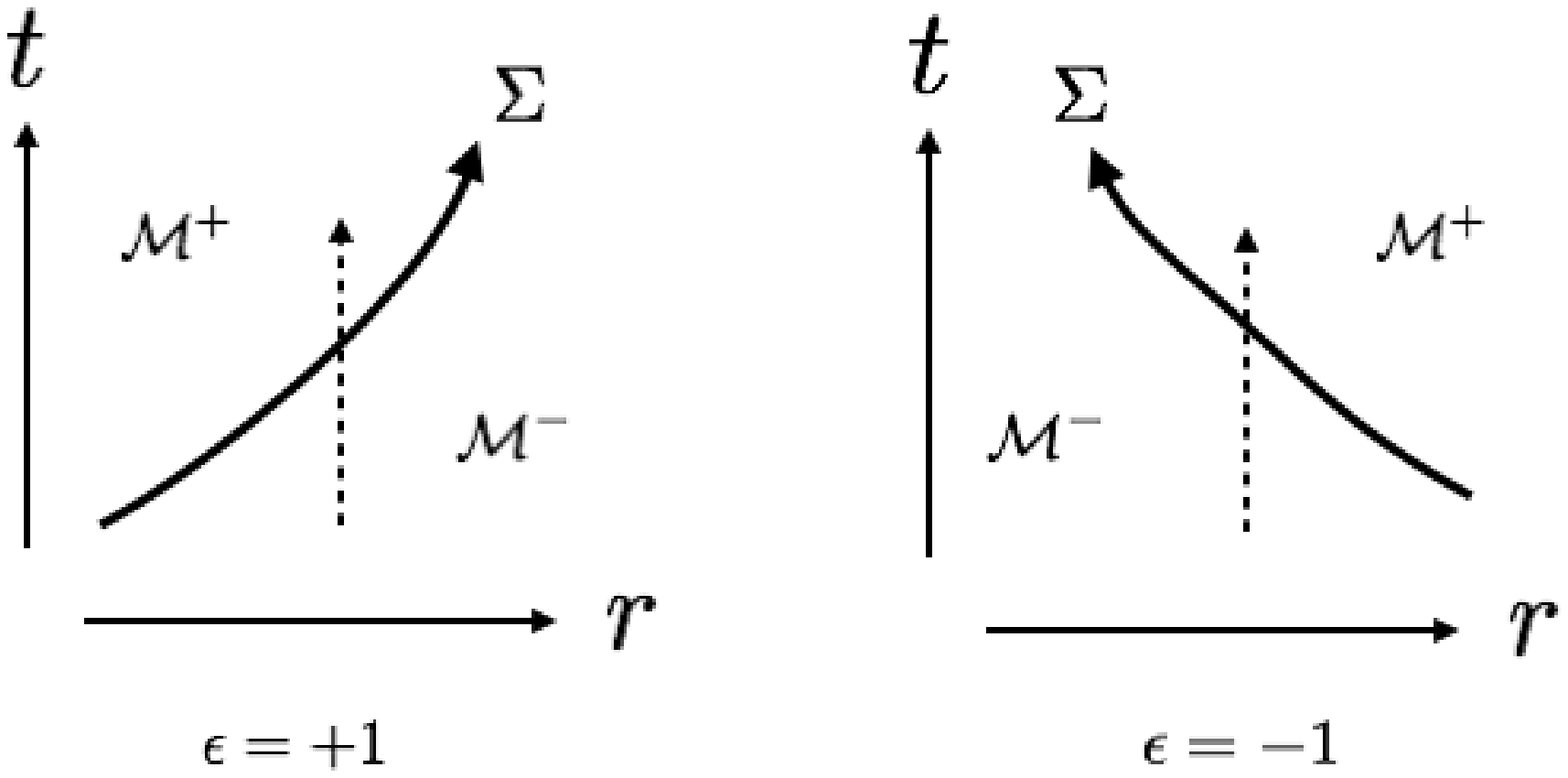}
\caption{ Schematic figure for null matching.  Null hypersurface
 $\Sigma$ partitions a spacetime into two region, past ($\mathcal M^-$)
 and future ($\mathcal M^+$) of $\Sigma$. The dashed line is the world line of an observer at fixed spatial coordinates. }
\label{fig-schematic-matching} 
\end{center}
\end{figure}

As we are interested in emission (absorption) from a naked singularity the two spacetimes are matched along a expanding (imploding) null surfaces. The equation describing such a surface is given by the outgoing (incoming) radial null geodesics in $\mathcal M^\pm$, {\it i.e.},
\begin{align}
\frac{\D t_\pm}{\D r_\pm}=\epsilon_\pm \sqrt{\frac{B_\pm}{|A_\pm|}}. \label{outgoing-null-geodesic}
\end{align}
Note that $A=-|A|$ for outside of the horizon (if any), and
$\epsilon_\pm$ is defined by
\begin{align}
\epsilon_+:=
\begin{cases}
    +1 & :{\rm outgoing} \\
    -1 & :{\rm incoming}
  \end{cases},~~
\epsilon_-:=
\begin{cases}
    +1 & :{\rm outgoing} \\
    -1 & :{\rm incoming}
  \end{cases}. 
\end{align}
The sign of $\epsilon$ represents null rays to be outgoing or incoming. One can choose a specific null geodesic in $\mathcal M^\pm$, that characterizes null hypersurfaces $\Sigma_+$ and $\Sigma_-$ in $\mathcal M^+$ and $\mathcal M^-$, respectively. The hypersurfaces $\Sigma_+$ and $\Sigma_-$ are identified so that the two spacetimes are glued along these null hypersurfaces, i.e., $\Sigma_+ = \Sigma_- = :\Sigma$. In this sense the null hypersurface $\Sigma$ partitions the spacetime into two regions; the past ($\mathcal M^-$) and the future ($\mathcal M^+$) of $\Sigma$ (\fig{fig-schematic-matching}).

The two metrics must be same on $\Sigma$, {\it i.e.}, $\D s_+^2|_\Sigma=\D s_-^2|_\Sigma$, and from Eq. (\ref{general-metric}), the coefficient of the angular part gives a relation between $C_+$ and $C_-$ :
\begin{align}
\left.C_+\left(t_+(r_+),r_+\right)\right|_{\Sigma}=\left.C_-\left(t_-(r_-),r_-\right)\right|_{\Sigma}. \label{relation-plus-minus-1}
\end{align}

\subsection{Null hypersurface - intrinsic coordinates}

To implement the null shell formalism we first consider a system of coordinates intrinsic to the hypersurface $\Sigma$: 
\[
y^l=(\lambda, \theta, \varphi) ,
\]
here $\lambda$ is an arbitrary parameter characterizing generators on the null hypersurface. The hypersurface can be described as $x^\alpha_\pm=(t_\pm(\lambda), r_\pm(\lambda), \theta, \varphi)$ in terms of coordinates on either side. Thus, the parametric equations for the null hypersurface $\Sigma$, as seen from either side, are:
\begin{align}
t_\pm=t_\pm(r_\pm(\lambda)),~~r_\pm=r_\pm(\lambda),~~\theta=\theta,~~\varphi=\varphi.
\end{align}
The vectors $k^\alpha$, tangent to the null hypersurface on each side are given by
\begin{align}
k^\alpha_\pm \partial_\alpha=
\left. \frac{\D t_\pm}{\D r_\pm}\right|_\Sigma \mathring r_\pm \partial_t+ \mathring r_\pm \partial_r,
\end{align}
where $\mathring{}:=\D/\D \lambda$. Because a radial null geodesic equation for $\Sigma$ is given by \eq{outgoing-null-geodesic},
$k^\alpha_\pm$ reduces to
\begin{align}
k^\alpha_\pm \partial_\alpha=\epsilon_\pm \sqrt{\frac{B_\pm}{|A_\pm|}}\mathring r_\pm \partial_t+ \mathring r_\pm \partial_r. \label{general-null-vector}
\end{align}
Vector fields $e^\alpha_{(\theta)}\partial_\alpha=\partial_\theta$ and $e^\alpha_{(\varphi)}\partial_\alpha=\partial_\varphi$ are also tangent to $\Sigma$. Moreover, auxiliary null vector fields $N^\alpha_\pm$ satisfying $N^\alpha_\pm k_{\alpha \pm}=-1, N^\alpha_\pm N_{\alpha \pm}=0$ and $N^\alpha_\pm e_{\alpha \pm}=0$ are obtained as
\begin{align}
N_\alpha^\pm \D x^\alpha=-\frac{\epsilon_\pm}{2\mathring r_\pm}\sqrt{\frac{|A_\pm|}{B_\pm}}\D t -\frac{1}{2\mathring r_\pm}\D r. \label{auxiliary-null-vector}
\end{align}
Vectors, $k^\alpha, e^\alpha_{(\theta)}, e^\alpha_{(\varphi)}$ and $N^\alpha$ complete a basis on $\Sigma$ \cite{toolkit}.

\subsection{Surface stress-energy tensor and transverse curvature}
 We calculate the surface stress-energy tensor, {\it i.e.}, a stress-energy tensor on $\Sigma$, which is written as
\begin{align}
T^{\alpha \beta}_\Sigma=(-k^\gamma u_\gamma)^{-1}S^{\alpha \beta}\delta(\tau),
\end{align}
where
\begin{align*}
    S^{\alpha \beta}:= \mu k^\alpha k^\beta+p\sigma^{AB}e^\alpha_{(A)}e^\beta_{(B)}.
\end{align*}
Here, $\mu$ and $p$ are interpreted as the surface energy density and the surface pressure on the null hypersurface, respectively. The parameter $\tau$ was introduced as the proper time of an observer who has the four velocity $u^\alpha$; it takes zero value while crossing $\Sigma$. Defining transverse curvature $C_{lm}$, as
\[
C_{lm}:=-N_\alpha e^\alpha_{(l);\beta } e^\beta_{(m)},
\] 
we can write the explicit expressions for the energy density $\mu$, and the pressure $p$, as: 
\begin{align}
8\pi\mu:=-\sigma^{ab}[C_{ab}], \qquad 8\pi p:=-[C_{\lambda \lambda}].
\end{align}
The indices $l,m$ run through $\lambda, \theta $ and $\varphi$, and $a, b$ through $\theta$ and $\phi$. The standard notation $[A]:=A_+|_\Sigma-A_-|_\Sigma$, is used for measuring jump in scalar qualities at $\Sigma$.

The general expressions for the non-zero components of transverse curvature can be calculated as (hereafter, $\pm$ sign is omitted for clarity):
\begin{align}
C_{\lambda \lambda}
=&\mathring r \left\{\epsilon \frac{\dot B}{\sqrt{|A|B}}+\frac{1}{2}(\ln |A|)'+\frac{1}{2}(\ln B)' \right\}
 \label{C-lambda-lambda}, \\
C_{ab}
&=-\frac{\sigma_{ab}}{4C \mathring r}\left(-\frac{\epsilon \dot C}{\sqrt{|A|B}}+\frac{C'}{B} \right), \label{C-AB}
\end{align}
where $\sigma_{ab}\D x^a \D x^b:=C(\D \theta^2+\sin^2\theta \D \varphi^2)$.

\subsection{Parameter independent description for the shell's whole
energy and relation to its luminosity}
We define the integrated energy over a spherical shell, which would be measured by an observer who is in $\mathcal M^-$ with a four velocity $u_\alpha^-$, as
\begin{align}
E_{{\rm shell}}:= 4 \pi \left. C T_{\alpha \beta} u^\alpha_- u^\beta_-\right|_{\Sigma}. \label{E-shell}
\end{align}
We should note that the quantity $E_{{\rm shell}}$ is actually related to a {\it luminosity} emitted from the system.  
Luminosity of the shell can be defined as the area integration of energy flux that would be measured by a comoving observer at infinity (which means he/she is a static observer), who has the four velocity $u^\alpha$. In equation this is written as 
\begin{align}
L_{{\rm shell}}:=4\pi \left. C T_{\alpha \beta}u^\alpha n^\beta \right|_{\Sigma}, \label{L-shell}
\end{align}
where $n^\beta$ is one of three spacelike unit basis vectors which are orthogonal to $u^\alpha$ and its component towards to the direction in which the shell moves: 
\begin{align}
n^\alpha \partial_\alpha=\partial_r, n^\alpha n_\alpha=1, u^\alpha n_\alpha=0. 
\end{align}

By geometrical observation, one identifies the null vector $k^\alpha$ is proportional to $u^\alpha+n^\alpha$: $k^\alpha=b(u^\alpha+n^\alpha)$ with $b=$constant. 
 Using all properties for $k^\alpha, u^\alpha$ and $n^\alpha$, $E_{{\rm shell}}$ and $L_{{\rm shell}}$ reduce to
\begin{align}
&E_{{\rm shell}}=4\pi C(-b)\mu, \\
&L_{{\rm shell}}=4\pi C(-b) \mu.
\end{align}
Thus, 
we have a relation
\begin{align}
E_{{\rm shell}}=L_{{\rm shell}}.
\end{align}

$E_{{\rm shell}}$ (or equivalently $L_{{\rm shell}}$) is derived by
junction conditions and thus this is functions of the both of the
metrics, $g_{\mu\nu}^+$ and $g_{\mu\nu}^-$. However, since we observe $E_{{\rm shell}}$ from the either side of the two spacetimes, the final description of physical quantities $E_{{\rm shell}}$ must be written in the metric in the one side. The two metrics are related each other through \eq{relation-plus-minus-1}.
We will denote $E_{{\rm shell}}$ by the metric in $-$ side. 
To do so,
consider \eq{relation-plus-minus-1}, it relates $r_+$ and $r_-$, say :
\begin{align}
r_+ = \psi(r_-). \label{relation-plus-minus-2}
\end{align}
Explicit functional form of $\psi$ can be determined once the line elements $\mathcal M^+$ and $\mathcal M^-$ are fixed. Since $r_\pm$ is parametrized by $\lambda$, differentiation of \eq{relation-plus-minus-2} gives
\begin{align}
\mathring r_+=\frac{\D \psi(r_-)}{\D r_-}\mathring r_-. \label{relation-plus-minus-deriv}
\end{align}
Using \eq{relation-plus-minus-2} and \eq{relation-plus-minus-deriv}, we can eliminate $r_+$ and $\mathring r_+$ from the expressions of $\mu$ and $p$. Consequently, the energy $E_{{\rm shell}}$ becomes a function of $r_-$ only. 

\section{Examples}\label{Examples}
 
\subsection{Energy emission from a naked singularity: self-similar dust collapse}\label{example1}
We consider naked singularity formation in self-similar dust collapse. We consider two spacetime manifolds $\mathcal M^+$ and $\mathcal M^-$ as two spherically symmetric dust spacetimes whose metrics are given by
\begin{align}
\D s_\pm^2=-\D t_\pm^2+(R'_\pm)^2 \D r_\pm^2+R_\pm^2(\D \theta^2+\sin^2 \theta \D \varphi^2), \label{dust-metric}
\end{align}
where $R_\pm=R_\pm(t_\pm,r_\pm)$. Or equivalently from \eq{general-metric}, we adopt the functions as follows:
\begin{align}
A_\pm=-1,~B_\pm=(R_\pm')^2,~C_\pm=R_\pm^2,~\epsilon_\pm=1.
\end{align}
Stress-energy tensor for dust is given by $T^{\alpha\beta}_\pm = \rho_\pm(t,r) u^\alpha_\pm u^\beta_\pm$, with co-moving four-velocity of dust $u^\alpha_\pm=\delta^\alpha_0$, and dust energy density$\rho_\pm(t,r)$. The non-trivial Einstein field equations derived from the above metric and stress-energy tensor yield ($\pm$ sign is omitted for brevity):
\begin{align}
& \dot R^2 = \frac{F(r)}{R}, \label{dot-R} \\
& F' = 8\pi \rho R^2 R', \label{deriv-F}
\end{align}
where $F(r)$ is an arbitrary function of $r$. 
Using the remaining scaling freedom we set $R(0,r)=:r$, and \eq{deriv-F} can now be integrated to fix $F$ in terms of initial density profile $\rho(0,r)$ as: 
\begin{align}
F(r)= 8 \pi \int \rho(0,r)r^2\D r.
\end{align}
Hence $F$ is twice the mass interior to a sphere of radius $r$.

\eqs{dot-R} and (\ref{deriv-F}) can be solved simultaneously for the unknown metric function $R$, 
\begin{align}
R^{3/2}=r^{3/2}-\frac{3}{2}\sqrt{F}t.
\end{align}
This solution is known as the marginally bound Lema\^itre-Tolman-Bondi (LTB) solution. 
If one takes linear mass function, {\it i.e.} $F = 2\kappa r$, we have (with recovering $\pm$ sign)
\begin{align}
R_\pm^{3/2} = r_\pm^{3/2} \left( 1 - a_\pm\frac{t_\pm}{r_\pm}\right), \label{function-R}
\end{align}
where $a_\pm:=3\sqrt{2\kappa_\pm}/2$, and $\kappa_\pm$ is assumed to be non negative constant. Minkowski spacetime is recovered for $\kappa_\pm=0$.
We also note here that the spacetime of \eq{dust-metric} with metric function of the form \eq{function-R}, admits self-similarity \cite{carr05}. To maintain simplicity and clarity in further discussion the attention  is restricted to the self-similar solution. However, with some difficulty, the discussion can be extended to the non-self-similar case also.  

In this well studied setup the singularity forms first at the symmetry
center $(t,r)=(0,0)$, and this can be globally naked for a range of
values of parameter $\kappa$, characterizing strength of gravity
\cite{Barve+1999}. Actual value of the parameter range is not important
and from now on the discussion assumes spacetime with the presence of a
naked singularity. Considering radial null geodesics emanating from the
singularity at the center, the  Cauchy horizon is defined as the fastest
null ray coming out of the singularity. To identify equation of the Cauchy
horizon, we first identify the family of singular outgoing radial null
geodesics. From \eq{dust-metric}, the outgoing radial null geodesics
obeys the equation $\D t/\D r=R'$ (we omit the  $\pm$ sign from expressions for brevity) which is explicitly written as 
\begin{align}
\frac{\D t}{\D r}=\frac{3 r - a t }{3r^{2/3}(r-at)^{1/3}}. \label{orange-explicit}
\end{align}
The condition for existence of out-going radial null rays, meeting central singularity $(t,r)=(0,0)$ in their past with a positive finite slope, can be reduced to a finite non-zero positive value of the following limit:
\begin{align}
\lim_{t \to 0, r \to 0} \frac{t}{r}=\lim_{t \to 0, r \to 0} \frac{\D t}{\D r}=z, \label{orange-limit}
\end{align}
where $z$ takes a constant value. The first equality in equation above holds due to l'H\^opital's rule. Then, \eq{orange-explicit} reduces to a  quartic equation (restoring the $\pm$ sign)
\begin{align}
f(z_\pm):=a_\pm z_\pm^4-\left(1+\frac{a_\pm^3}{27}\right)z_\pm^3+\frac{a_\pm^2}{3}z_\pm^2-a_\pm z_\pm+1=0 . \label{quartic-z}
\end{align}
Using properties of quartic equations, in \eq{quartic-z} some information on the nature of roots can be determined from discriminant $D=(-4a_\pm^6+ 2808a_\pm^3-729)/27$. If $D < 0$, the quartic has two real roots, and for $D=0$, there is one real root. $D$ is negative when  $0\leq a_\pm < a_*:=3/(2(26+15\sqrt{3}))^{1/3}=0.638014$ and vanishes when $a_\pm=a_*$
\footnote{In fact, the condition $D<0$ gives two possible range for $a$, $0\leq a<a_*=0.638$ or $a>a_*=8.886$. However, the larger range of $a_*$ is unphysical because outgoing radial null geodesics form after the shell focusing singularity \cite{Duffy-Nolan2011}.}. The range of $a_\pm$ implies $0\leq \kappa_\pm \leq \kappa_*:=0.0904583$. 
In the self similar dust solution, \eq{dust-metric} and \eq{function-R}, the Cauchy horizon takes the form:
\begin{align}
t_\pm=z_{ch}^\pm r_\pm, \label{CH}
\end{align}
where $z_{ch}^+$ and $z_{ch}^-$ are the positive constants in $\cal M^+$ and $\cal M^-$, respectively. These are the smallest positive values of the solution to \eq{quartic-z}.

We consider matching two self similar dust spacetimes at their Cauchy horizons which are given by \eq{CH}. So, the Cauchy surface is that null hypersurface $\Sigma$ which partitions the spacetime into two regions; the past and the future of $\Sigma$. The matching procedure imposes restrictions on the metric function and its first derivative, respectively, the first and the second fundamental forms. 

We start with identification of outgoing null geodesics in two spacetimes. Recall, from \eq{dust-metric}, the outgoing radial null geodesics are
\begin{align}
\frac{\D t_\pm}{\D r_\pm}=R_\pm'. \label{orange}
\end{align}
The matching of first fundamental form, equivalent to condition \eq{relation-plus-minus-1}, implies that metrics are same on $\Sigma$, yielding $R_+|_\Sigma=R_-|_\Sigma$, i.e.,
\begin{align}
\left(1-a_+z_{ch}^+\right)^{2/3}r_+|_\Sigma
=\left(1-a_-z_{ch}^-\right)^{2/3}r_-|_\Sigma. \label{1st-fundamental}
\end{align}
Here, we have used the equation $t_\pm/r_\pm=z_{ch}^\pm$ on $\Sigma$. \eq{1st-fundamental} is our first fundamental form and this gives a relation between $r_+$ and $r_-$ on $\Sigma$.

The null hypersurface in parametric form can be written as:
\begin{align}
t_\pm=z_{ch}^\pm r_\pm(\lambda),~~r_\pm=r_\pm(\lambda),~~\theta=\theta,~~\varphi=\varphi.
\end{align}
We construct now the explicit set of basis vectors for the self-similar case under consideration to implement the null-shell formalism.
From \eq{general-null-vector} the tangent null vector $k^\alpha_\pm$ is given by
\begin{align}
k^\alpha_\pm \partial_\alpha= z_{ch}^\pm \mathring{r}_\pm \partial_t + \mathring{r}_\pm \partial_r,\label{null-vector1}
\end{align}
The spatial unit tangents to sphere are given by $e^{\alpha~\pm}_{(\theta)}\partial_\alpha=\partial_\theta$ and $e^{\alpha~\pm}_{(\varphi)}\partial_\alpha=\partial_\varphi$.
The  auxiliary null vector $N^\alpha_\pm$ are written by \eq{auxiliary-null-vector} as
\begin{align}
N^\alpha_\pm \partial_\alpha=-\frac{1}{2z_{ch}^\pm \mathring{r}_\pm}\partial_t -\frac{1}{2(z_{ch}^{\pm})^2 \mathring{r}_\pm}\partial_r. \label{null-vector2}
\end{align}
Now that we have explicit expressions for the basis vectors above, we have the tools to implement the junction conditions for matching two null hypersurfaces. The non-zero components of the transverse curvature, \eq{C-lambda-lambda} and \eq{C-AB}, are
\begin{align}
&C^\pm_{\lambda \lambda}
=\mathring{r}_\pm \frac{2a_\pm^2z^\pm_{ch}}{9r_\pm \left(1-a_\pm z^\pm_{ch}\right)^{4/3}},\\
&C_{ab}^\pm
=-\frac{1}{\mathring{r}_\pm}\frac{\sigma_{ab}}{2z^\pm_{ch}R_\Sigma^\pm}\left(\frac{2a_\pm}{3(1-a_\pm z^\pm_{ch})^{1/3}}+1 \right),
\end{align}
where $\sigma_{ab}\D x^a\D x^b:=R^2(\D \theta^2+\sin^2 \theta \D \varphi^2)$.

Consequently, $\mu$ and $p$ can be written as
\begin{align}
8\pi \mu =&\frac{1}{z^+_{ch}R_\Sigma^+ \mathring{r}_+}\left(\frac{2a_+}{3(1-a_+ z^+_{ch})^{1/3}}+1 \right)
-\frac{1}{z^-_{ch}R_\Sigma^- \mathring{r}_-}\left(\frac{2a_-}{3(1-a_- z^-_{ch})^{1/3}}+1 \right),  \label{surface-energy} \\
8\pi p=&-\mathring{r}_+ \frac{2a_+^2z^+_{ch}}{9r_+ \left(1-a_+ z^+_{ch}\right)^{4/3}}
+\mathring{r}_- \frac{2a_-^2z^-_{ch}}{9r_- \left(1-a_- z^-_{ch}\right)^{4/3}}.
\label{surface-pressure}
\end{align}
As can be seen from expressions \eq{surface-energy} and \eq{surface-pressure}, the surface energy and the surface pressure are determined by fixing initial data, namely, the parameter $\kappa_+$ and $\kappa_-$. The $\mathring{r}_+$,  in equations above, is related to $\mathring{r}_-$  through \eq{1st-fundamental}. 
Clearly, no surface energy and pressure are carried if $a_+=a_-$ (or equivalently $\kappa_+  = \kappa_-$), i.e., the initial profile $\kappa$ is same in $\mathcal M^+$ and $\mathcal M^-$. If $a_{-} < a_{+}$, $E_{shell} > 0$, and for $E_{shell} < 0$, $a_{-} > a_{+}$.

\begin{figure}[thb]
\begin{center}
\includegraphics[clip,width=10cm]{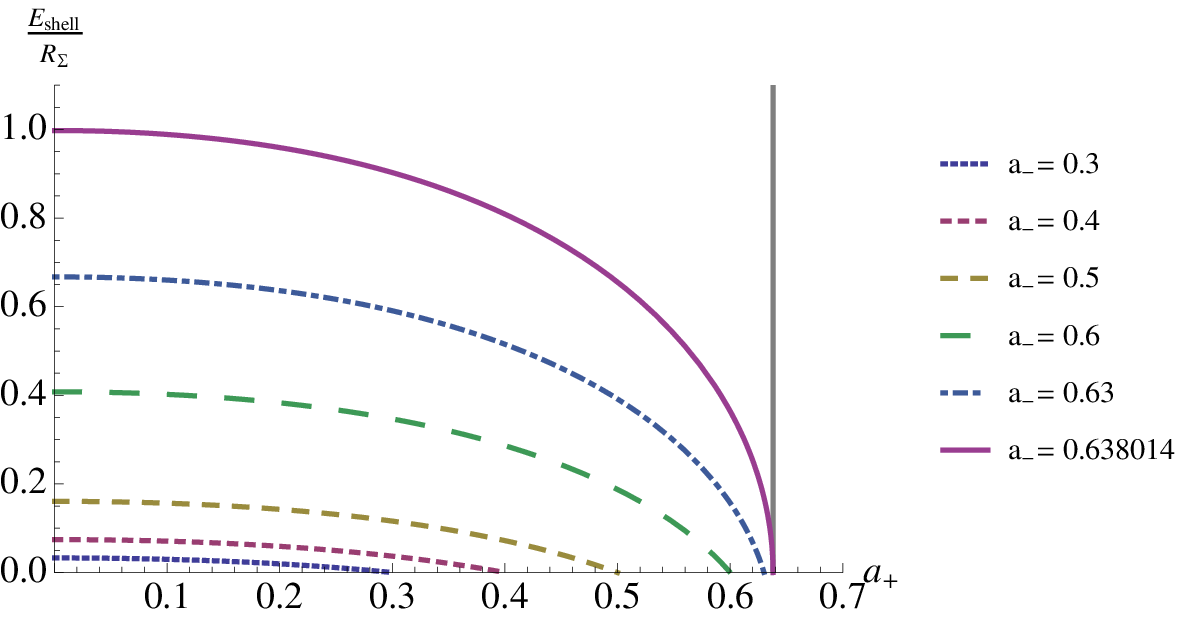}
\caption{$E_{{\rm shell}}/R_{\Sigma}$ versus $a_+$. The vertical solid
 line is at $a_+=a_*=0.638014$.
$E_{{\rm shell}}/R_{\Sigma}$ is positive only for $0\le a_{+}< a_{-}\le a_{*}$. The more the gap between $a_-$ and $a_+$ increases, the more $E_{{\rm shell}}/R_{\Sigma}$. 
}
\label{fig-example1-ap-am} 
\end{center}
\end{figure}

From \eq{E-shell}, the energy of the null shell $E_{{\rm shell}}$ that is measured by an observer in the past of $\Sigma$ (we see $\Sigma$ from the $(-)$ side) with the four velocity $u^\alpha_-$:
\begin{align}
E_{{\rm shell}}  = \left.4\pi R^2T^{\alpha \beta} u^-_\alpha u^-_\beta \right|_{\tau=0}. \label{Eshell}
\end{align}
For a comoving observer choosing the four velocity to be $u^\alpha_-\partial_\alpha=\partial_t$,  the expression for energy \eq{Eshell} takes the form
\begin{align}
E_{{\rm shell}}=4\pi R^2_{\Sigma}(-k^\gamma_- u^-_\gamma) \mu. \label{Eshell-comoving}
\end{align}
One calculates $-k^\gamma_- u^-_\gamma=\mathring{r}_-R'_-$ and hence \eq{Eshell-comoving} reduces to
\begin{align}
E_{{\rm shell}}
= \frac{R_{\Sigma}}{2}\left\{ \frac{\left(1-a_+z_{ch}^+\right)^{2/3}z_{ch}^-}{\left(1-a_-z_{ch}^-\right)^{2/3}z_{ch}^+}\left(\frac{2a_+}{3}\left(1-a_+z_{ch}^+\right)^{-1/3}+1\right)
-\left(\frac{2a_-}{3}\left(1-a_-z_{ch}^-\right)^{-1/3}+1\right) \right\}. \label{E-shell-dust-dust}
\end{align}
One can see that $E_{{\rm shell}}$ is linear function of $R_{\Sigma}$.
Numerical plot for $E_{{\rm shell}}/R_{\Sigma}$  in
\eq{E-shell-dust-dust} is shown in
\fig{fig-example1-ap-am}.

The \fig{fig-example1-ap-am} describes variation of the ratio $E_{{\rm
shell}}/R_{\Sigma}$ with $a_+$, the energy density parameter of the
matched spacetime, for fixed values of $a_-$. The parameters, $a_+$ and
$a_-$, take values in the following range, $0\le a_{+}\le a_{-}\le a_{*}$. 
For a fixed value of $a_-$, the ratio $E/R$ decreases with increase in $a_+$, and vanishes for $a_+ = a_-$ due to positivity of energy emitted. The maximum emission for a given $a_-$ lies at $a_+ =0$, which corresponds to Minkowski spacetime in the matched region. Physically this implies all the energy in $M_-$ is emitted along the null-shell. For the largest allowed value of $a_- = a_*$, beyond which $M_-$ evolves into a black hole spacetime geometry, $E_{{\rm shell}}/R_{\Sigma}$ is the largest for $a_+= 0$.

Both in the dust case (the model given in the present paper) and in the Vaidya case (Jhingan {\it et al.}'s model), the surface energy $\mu$ is proportional to $r^{-1}$. This fact implies that the whole energy of the null shell, $E_{{\rm shell}}$, grows in proportion to $r$ (or equivalently $R_\Sigma$) and hence diverges at spatial infinity. This blow-up feature is essentially due to the self-similarity of spacetimes. The reason is following: regardless of whether spacetime is self-similar or not, the surface energy $\mu$ has a dimension of $1/({\rm length})$. When we restrict our attention to self-similar spacetime, then the description of $\mu$ must be of the form $F(z)/r$, where $z:=t/r$. Because $z$ is constant on the Cauchy horizon (and also on the event horizon) in self-similar metric, $F$ becomes $F(z_{ch})$ which is definitely constant. Thus, $E_{{\rm shell}} \propto 4\pi (R|_\Sigma)^2 \mu$ (where $R|_\Sigma=r|_\Sigma\left(1-az_{ch}\right)^{2/3}$ in our dust model and $R|_\Sigma=r|_\Sigma$ in Vaidya model) must be proportional to $r$ (or equivalently $R_\Sigma$) under the self-similar assumption.

We give a physical interpretation of the increasing energy of the null shell. Take the conservation law for null shell which is given by Poisson's book \cite{toolkit} as
\begin{align}
p\frac{\D}{\D \lambda}\delta S+[T_{\alpha\beta}k^\alpha k^\beta]\delta S=0,
\label{cons-null-shell}
\end{align}
where $\delta S:= C \sin^2\theta \D \theta \D \varphi$ is an element of cross-sectional area on the shell. 
The first term on the left-hand side of \eq{cons-null-shell} is the work done by the shell's expansion, while the second term is the energy absorbed into the shell from its surrounding. 
Physical interpretation for growing of the shell energy $4\pi R^2 \mu$ is that; the null shell expands in an imploding dust region while absorbing the energy of the surrounding dust.

\subsubsection{Instant singularity}
As an interesting example of our model, one can consider an exclusive situation, say, ``a singularity in an instant": a singularity disappears immediately after its occurrence. To describe this, let us take $0<\kappa_- \leq \kappa_*$  in the past of $\Sigma$. Then a singularity forms in a finite time.
In the $(+)$ side, we take $\kappa_+=0$ $(a_+=0)$, the minimum value of $\kappa$. Vanishing $\kappa$ represents the Minkowski spacetime.
Since $\mathcal M^+$ is defined as the future of $\Sigma$, the above setup describes an instant presence of a singularity. The singularity exists only at the point $(t,r)=(0,0)$.
 For $a=0$, we have $z_{ch}=1$ from  \eq{quartic-z}. Thus, in this scenario, the energy of the shell that appeared from the point $(t,r)=(0,0)$ reduces to
\begin{align}
E_{{\rm shell}}= \frac{R_{\Sigma}}{6}\left\{ \frac{3z_{ch}^-}{\left(1-a_-z_{ch}^-\right)^{2/3}}
-\left(2a_-\left(1-a_-z_{ch}^-\right)^{-1/3}+3\right) \right\}.
\label{surface-energy-no-singularity}
\end{align} 
For simplicity, let us take $a_-=a_* \Leftrightarrow \kappa_-=\kappa_*$, the maximum value of $\kappa$.
For $a=a_*=0.638014$, $z_{ch}$ is solved as $z_{ch}(a_*)=1.25992$. By putting these values into  \eq{surface-energy-no-singularity}, we have $E_{{\rm shell}}=R_{\Sigma}$.

According to this scenario, a region of the dust matter is pushed away from the center by the motion of outgoing null shell.

\subsection{Energy emission from a naked singularity: Negative mass Schwarzschild solution}\label{example2}
We have considered an isotropic emission of radiation from a naked singularity by modeling two self-similar LTB spacetimes with different density profiles in the first example. Here, we consider another emission scenario which describes a null emission from a singularity after dust collapse in which the final singularity has negative mass. This situation is modeled by taking $\mathcal{M}^-$ as LTB solution while $\mathcal{M}^+$ is a negative mass Schwarzschild solution:
\begin{align}
&A_-=-1,~B_-=(R'(t_+,r_+))^2,~C_-=R(t_+,r_+)^2, \nonumber \\
&A_+=-f,~B_+=f^{-1},~C_+=r_+^2,~f:=1+\frac{2|M|}{r_+}, ~\epsilon_\pm=1,
\end{align}
where $M$ is a negative constant.

In $\mathcal M^-$, the fastest outgoing null geodesic emanating from the center $(t_-,r_-)=(0,0)$ makes the Cauchy surface and it is same equation in the previous example: $t_-=z_{ch}r_-$. $z_{ch}$ is the smallest solution to \eq{quartic-z}. Also, other quantities relevant to our calculation are same.

In $\mathcal M^+$, the outgoing null geodesic is given by
\begin{align}
\frac{\D t_+}{\D r_+}=\frac{r_+}{r_++2|M|}. \label{radial-sch}
\end{align}

Especially, the null emitted from the center is explicitly given by integrating \eq{radial-sch}:
\begin{align}
t_+=r_+-2|M|\ln \left(\frac{r_+}{2|M|}+1 \right).
\end{align}
The condition for the both metrics to be same at $\Sigma$, \eq{relation-plus-minus-1}, takes the form $R_-|_\Sigma=r_+|_\Sigma$, {\it i.e.},
\begin{align}
\left(1-az_{ch}\right)^{2/3}r_-(\lambda)|_\Sigma=r_+(\lambda)|_\Sigma.
\end{align}
The parametric equation in $\mathcal M^+$ is 
\begin{align}
t_+=r_+(\lambda)-2|M|\ln \left(\frac{r_+(\lambda)}{2|M|}+1 \right),~~r_+=r_+(\lambda),~~\theta=\theta,~~\varphi=\varphi.
\end{align}
Basis vectors on $\Sigma$ are
\begin{align}
k_+^\alpha \partial_\alpha=\mathring r_+ f^{-1}\partial_t+\mathring r_+ \partial_r,~~
N_\alpha^+ \D x^\alpha=-\frac{f}{2\mathring r_+}\D t_+ -\frac{1}{2\mathring r_+}\D r_+,~~
e^\alpha_{(\theta)} \partial_\alpha=\partial_\theta,~~e^\alpha_{(\varphi)} \partial_\alpha=\partial_\varphi.
\end{align}
Transverse curvature $C^+_{lm}$ is calculated as
\begin{align}
C_{\lambda \lambda}^+=0,~~
C^+_{ab}=-\frac{f}{2r_+\mathring r_+}\sigma^+_{ab},
\end{align}
where $\sigma^+_{ab} \D x^a \D x^b:=r_+^2(\D \theta^2+\sin^2\theta \D \varphi^2)$.
Then we have $\mu$ and $p$ as
\begin{align}
8\pi \mu=&\frac{f}{r_+\mathring r_+}-\frac{1}{z_{ch}R_\Sigma \mathring{r}_-}\left(\frac{2a}{3(1-az_{ch})^{1/3}}+1 \right), \\
8\pi p=& \mathring{r}_- \frac{2a^2z_{ch}}{9r_- \left(1-a z_{ch}\right)^{4/3}}.
\end{align}

For calculation of $E_{{\rm shell}}$, as same as in the previous example, we adopt the four velocity of a comoving observer. Then the final form for the shell energy is written by 
\begin{align}
E_{{\rm shell}}=\frac{z_{ch}|M|}{(1-az_{ch})^{2/3}}+\frac{R_\Sigma}{2}\left\{\frac{z_{ch}}{(1-az_{ch})^{2/3}}-\frac{2a}{3(1-az_{ch})^{1/3}}-1  \right\}. \label{E-shell-dust-sch}
\end{align}
In \eq{E-shell-dust-sch}, the first term is constant while the second term is linear in terms of $R_\Sigma$.
Because the coefficient of the second term is positive \cite{positivity-proof}, the energy increases monotonically as the null surface with $R_\Sigma$ expands.

Comparing \eq{E-shell-dust-sch} with the previous case, \eq{E-shell-dust-dust}, the first term on the right does not have $R_{\Sigma}$ dependence. This term corresponds to the energy distribution of the negative mass Schwarzschild solution that is point like. Note here that in this example the null-shell carries energy of the collapsing dust matter as well as energy relevant to the remnant negative mass.

In the next subsection Sec. \ref{example3}, we discuss appearance of a
negative mass Schwarzschild solution (naked singularity) from a
Minkowski spacetime by spontaneous emission of radiation. We can recover
this case in the $a \rightarrow 0$ limit of \eq{E-shell-dust-sch}. In
this limit we have $z_{ch} \rightarrow 1 $, from \eq{quartic-z}, and $E_{shell} \to |M|$.

\begin{figure}[t]
\begin{center}
\includegraphics[clip,width=10cm]{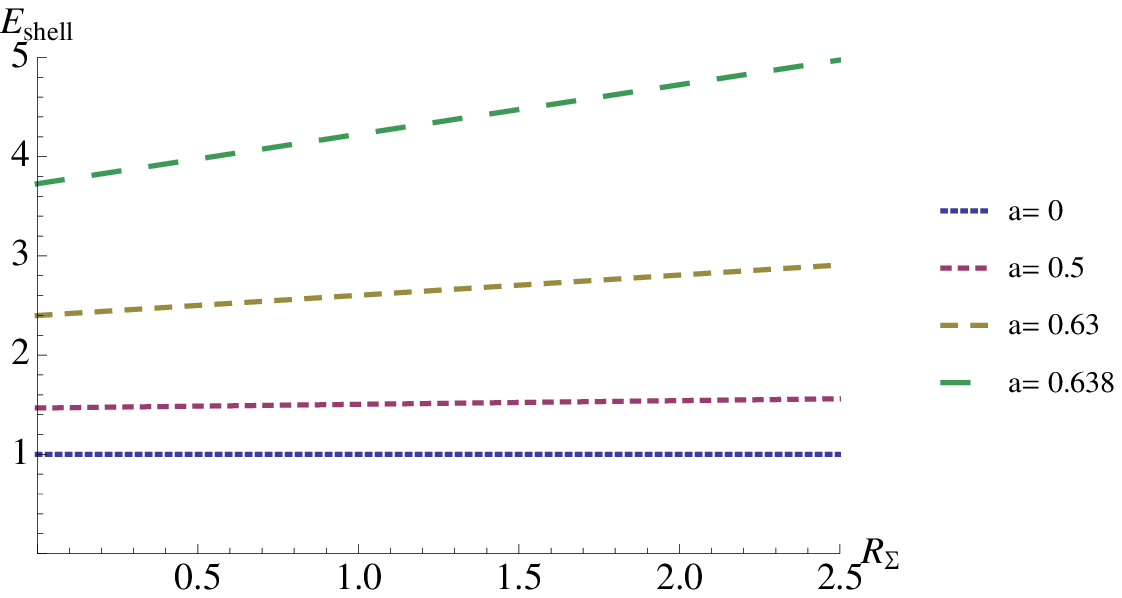}
\caption{ Values of $E_{{\rm shell}}$ in \eq{E-shell-dust-sch} for each $a$. We fixed $M$, say, $|M|=1$.
}
\label{fig-example2} 
\end{center}
\end{figure}

Numerical plot of typical parameters for $E_{{\rm shell}}$ in \eq{E-shell-dust-sch} is shown in \fig{fig-example2}.

In \fig{fig-example2}, energy of the shell is plotted for different value of initial density parameters.
The mass of the negative of Schwarzschild solution can be arbitrarily. The first curve ($a=0$) corresponds to Minkowski spacetime matching to negative mass Schwarzschild solution. As expected, the energy carried by the shell is equal to $|M|$.
The subsequent curves have additional energy corresponding to the LTB of the collapsing dust.

The situation in this example is similar to the previous example but the
future of $\Sigma$ is replaced with negative-mass Schwarzschild
solution. By operating junction conditions, we get \eq{E-shell-dust-sch}
for the energy of the expanding null shell. As in the previous example,
the shell's whole energy is a monotonically increasing function of $R$ and it diverges at infinity. 
At the same time the singularity forms, the null shell expands outward as the previous case does. However, the remnant singularity is much more serious than that of previous example due to the negative gravitational mass. 
Instant singularity, a special case  mentioned earlier can also be recovered in this example by taking the parameter $M=0$. 

\color{black}

\if.
\begin{figure}[htb]
\begin{center}
\includegraphics[clip,width=13cm]{fig-example2-a-M.eps}
\caption{ $E_{{\rm shell}}/R_{\Sigma}$ versus $a$. The dashed line is at $a=a_*$. Larger values of $M$ and/or $a$ make $E_{{\rm shell}}/R_{\Sigma}$ large.
}
\label{fig-example2-a-M} 
\end{center}
\end{figure}
\fi

\subsection{Spontaneous decay of Minkowski, end of inflation and black hole formation}\label{example3}
We consider several scenarios which denote spontaneous decay of Minkowski, end of inflation and  black hole formation. 
We can describe all of such scenarios under Schwarzschild-(A)dS solution in both spacetimes:
\begin{align}
A_\pm=-f_\pm,~ B_\pm=f_\pm^{-1},~ C_\pm=r_\pm^2,~ f_\pm:=1-\frac{2m_\pm}{r_\pm}-\frac{\Lambda_\pm}{3}r_\pm^2, ~\epsilon_+=\epsilon_-=:\epsilon.
\end{align}
The condition that the two metrics are same on null hypersurface, \eq{relation-plus-minus-1}, now yields just a simple relation,
\begin{align}
\left.r_+(\lambda)\right|_{\Sigma}=\left.r_-(\lambda)\right|_{\Sigma},
\end{align}
and consequently $\mathring r_+=\mathring r_-$.
$k^\alpha_\pm$ and $N_\alpha^\pm$ are given by
\begin{align}
& k_\pm^\alpha \partial_{\alpha \pm}=\epsilon_\pm \mathring r_\pm f_\pm^{-1}\partial_t^\pm+\mathring r_\pm 
\partial_r^\pm, \\
& N_\alpha^\pm \D x^\alpha=-\frac{\epsilon_\pm f_\pm}{2\mathring r_\pm}\D t -\frac{1}{2\mathring r_\pm}\D r. 
\end{align}
Transverse curvature on the both sides are now given by
\begin{align}
C^\pm_{\lambda\lambda}=0~~ {\rm and}~~
C^\pm_{ab}=-\frac{f_\pm \sigma^\pm_{ab}}{2r_\pm \mathring r_\pm}.
\end{align}
$\mu$ and $p$ are  written as
\begin{align}
&8\pi \mu=\frac{1}{r \mathring r}\left\{\frac{2}{r}(m_--m_+)+\frac{r^2}{3}(\Lambda_--\Lambda_+) \right\}, \\
&p=0,
\end{align}
where we defined $r:=\left.r_+(\lambda)\right|_{\Sigma}=\left.r_-(\lambda)\right|_{\Sigma}$.
When one adopt an observer who sticks in the coordinates in $\mathcal M^-$, a four-velocity takes the form $u_-^\alpha \partial_\alpha=\partial_t$ and hence $E_{{\rm shell}}$ can be evaluated as
\begin{align}
E_{{\rm shell}}
= \epsilon\left\{ (m_--m_+)+\frac{r^3}{6}(\Lambda_--\Lambda_+) \right\}. \label{E-shell-SAdS}
\end{align}
\eq{E-shell-SAdS} describes general result of shell energy seen by a distant observer.
We find \eq{E-shell-SAdS} is quartic function of $r$. If the gap of the cosmological constant is zero, $[\Lambda]=0$, the distant observer will measure a constant energy (luminosity). Otherwise, the energy is a function of $r^3$ which diverges at infinity.

Below, we investigate the models of a spontaneous decay of Minkowski, end of inflation and black hole formation as a consequence of contraction/expansion of null shell. We consider these scenarios as special cases of \eq{E-shell-SAdS}.

\subsubsection{Spontaneous decay of Minkowski}
We consider a case of  Minkowski solution in $\mathcal M^-$ and negative-mass Schwarzschild solution in $\mathcal M^+$. Then we take parameters as follows: 
\begin{align}
m_-=0, m_+=-|m_+|, \Lambda_\pm=0, \epsilon=+1.
\end{align}
This choice of geometries describes a spontaneous decay of Minkowski, a sudden appearance of a naked singularity which has a negative mass. In this case the null energy emitted from the singularity is evaluated as
\begin{align}
 \mu=\frac{|m_+|}{4\pi r^2\mathring r}, \quad E_{{\rm shell}}=|m_+|.
\end{align}
Since a negative-mass Schwarzschild spacetime describes a naked singularity, an appearance of a naked singularity occurs in a flat spacetime after null explosion having a positive energy.

One can consider another spontaneous decay model which shows more violent behavior than the previous one.
Suppose that null shell starts to expand from the center in Minkowski spacetime and also suppose that inside of the null shell is Anti de Sitter (AdS) spacetime. Then, this is the situation that also describes a spontaneous decay of Minkowski, instability of Minkowski spacetime. In this scenario the past of $\Sigma$, {\it i.e.}, $\mathcal M^-$, is Minkowski and the future of $\Sigma$, {\it i.e.}, $\mathcal M^+$, is AdS spacetime:
\begin{align}
m_\pm=0,  \Lambda_-=0, \Lambda_+=-|\Lambda|, \epsilon=+1.
\end{align}
Then we have
\begin{align}
\mu=\frac{|\Lambda|r}{24\pi \mathring r}, \qquad E_{{\rm shell}}=\frac{|\Lambda|}{6}r^3 
\label{decay-Minkowski-AdS}
\end{align}
which is  monotonically increasing function of $r$.

\subsubsection{End of inflation}
Let us take  Minkowski solution in $\mathcal M^-$ and de Sitter (dS) solution in $\mathcal M^+$.
This case is an example of the end of inflation and realized by taking parameters as follows: 
\begin{align}
m_\pm=0,  \Lambda_+=0, \Lambda_->0, \epsilon=+1.
\end{align}
Then the energy is given by
\begin{align}
\mu=\frac{\Lambda_-r}{24\pi \mathring r}, \qquad E_{{\rm shell}}=\frac{\Lambda_-}{6}r^3.
\label{end-inflation-dS-Minkowski}
\end{align}

Another example of the end of inflation is given by taking the parameters as 
\begin{align}
m_\pm=0, \Lambda_->\Lambda_+,  \epsilon=+1.
\end{align}
Then the energy is given by
\begin{align}
 \mu=\frac{(\Lambda_--\Lambda_+)r}{24\pi \mathring r}, \qquad E_{{\rm shell}}=\frac{(\Lambda_--\Lambda_+)}{6}r^3.
\label{end-inflation-dS-dS}
\end{align}

\subsubsection{Black hole formation}
An imploding null shell can result in black hole formation. We take such
an example by considering the choice of Minkowski solution in $\mathcal M^-$
and positive-mass Schwarzschild solution in $\mathcal M^+$, {\it i.e.},
\begin{align}
m_-=0, m_+>0,  \Lambda_\pm=0, \epsilon=-1.
\end{align}
Then the energy is given by
\begin{align}
\mu=\frac{m_+}{4\pi r^2 (-\mathring r)}, \qquad E_{{\rm shell}}=m_+
\end{align}
which is positive definite.
Here $\mu$ is positive because of shrink of the shell, i.e., $\mathring{r}<0$.

\if.
\color{red}
\subsubsection{Start of inflation?}
Start of inflation may be realized in a scenario of Minkowski solution in $\mathcal M^-$ and dS solution in $\mathcal M^+$. Null shell implodes toward the center of flat spacetime. 
\begin{align}
m_\pm=0,  \Lambda_-=0, \Lambda_+>0, \epsilon=-1.
\end{align}
Then the energy is given by
\begin{align}
 \mu=-\frac{\Lambda_+r}{24\pi \mathring r}, \qquad E_{{\rm shell}}=\frac{\Lambda_+}{6}r^3.
\end{align}

INTERPRETATION????
\fi

\subsubsection{Bubble nucleation}
One can consider phenomenological model of nucleation of bubble which is caused by transition of false vacuum.
Let us take parameters as follows:
\begin{align}
m_\pm=0,  \Lambda_->\Lambda_+, \epsilon=+1.
\end{align}
Then the energy is given by
\begin{align}
 \mu=\frac{(\Lambda_--\Lambda_+)r}{24\pi \mathring r}, \qquad E_{{\rm shell}}=\frac{(\Lambda_--\Lambda_+)}{6}r^3.
\label{bubble}
\end{align}

In our case the bubble propagates as the speed of light. The spacetime
before the shell passes is de Sitter, and the spacetime after passing
through the shell can be de Sitter ($\Lambda_{+}>0$), Minkowski ($\Lambda_{+}=0$), and AdS ($\Lambda_{-}<0$).
\section{Conclusion}\label{Conclusion}

We developed models of a radiation emitted from regions with extremely high curvature.  For this purpose, we derived a general formula which describes an imploding or exploding null shell  in general spherically symmetric spacetime. The energy is defined and evaluated as a surface energy of a null hypersurface, the null shell.

In general relativity, since one does not have predictability on
the evolution of a spacetime after a Cauchy horizon is formed, the future
of spacetimes made by matching on Cauchy horizons is not a unique
specification. Thus, we have investigated examples of filling the spacetime
inside of the Cauchy horizon and showed possibilities on the future
evolution of a singularity formation. We have proposed several models that
can describe dynamical processes of radiating energy, followed by the
gravitational collapse of a star.

In the first example, we constructed a model of energy emission by matching two self-similar marginally bound LTB spacetimes at their Cauchy horizons. Since the equation of the Cauchy horizon in the self-similar dust is well known, we can operate the explicit calculation for the matching. Our model describes an isotropic energy emission from the singularity. Due to the way of the construction of this model, a null shell starts expanding from the position of the singularity and carries certain energy which propagates along the Cauchy horizon. If we assume positive definiteness of the propagating energy, a condition $a_- > a_+$ must hold. We also derived an equation for the surface pressure which is defined on $\Sigma$. We found the surface energy $\mu$ is caused by the difference between the initial profiles $\kappa_+$ and $\kappa_-$ in the two dust spacetimes.  Such structure is qualitatively same as a model presented by Jhingan {\it et al.} \cite{Jhingan-Dwivedi-Barve2010}. They analyzed a model of matching the two Vaidya spacetimes at their Cauchy horizons and found the surface energy is obtained as the difference between the mass function in the $(+)$ side and in $(-)$ side.

We also proposed various examples focused on static spherically symmetric models. 
In most examples, $E_{{\rm shell}}$, or equivalently the absolute value of luminosity, is proportional to power of $r$. 
It is clear from \eq{E-shell-SAdS} that luminosity is constant if and only if the both cosmological constants are identical. Thus, increasing property of luminosity in Eqs. (\ref{decay-Minkowski-AdS}), (\ref{end-inflation-dS-Minkowski}), (\ref{end-inflation-dS-dS}) and (\ref{bubble}) is caused by the presence of difference between $\Lambda_+$ and  $\Lambda_-$. 
Since the cosmological constant can be interpreted as cosmological fluid, such increasing shell's energy should be supplied by the fluid, which is similar to the previous example treating dust spacetime in which increasing luminosity is caused by infalling dust fluid. 

Constant energy or luminosity, such as in the case of spontaneous decay of Minkowski, occurs only if $\Lambda_{+}=\Lambda_{-}$. 
Since there is no energy-supply, constant energy is taken to be the pure energy emitted from the central singularity.

We stress that the origin of emitted energy $E_{{\rm shell}}$ is divided
into two parts: energy itself from singularity and energy supply from
the fluid around the shell. 
Expanding null shell must increase its energy by absorbing energy of the neighboring fluid. 
In some parts of the paper we investigated behavior of expanding null shells in spacetime filled with fluid (self-similar LTB or cosmological fluid). In such situations, dust/cosmological fluid spreads to spacial infinity as simple example in calculation. In this case, due to the constant energy-supply to the shell from the fluid, a distant observer would measure violently energetic and luminous shell.
On the other hand, shell propagates having constant energy/luminosity in the case without fluid.

Lastly, we note that in astrophysical situation a fluid spreads to a
certain radius, so the energy-supply is cut off at the radius and after the null shell passes through that radius the shell's energy/luminosity should become a constant. 
Thus, such energy emission models created from higher curvature regions could be one of possible candidates playing role of high-energy phenomena in the universe.

\section*{Acknowlegements}
T. K. thanks Ken-ichi Nakao for stimulating discussions.
The authors wish to show thank to Vitor Cardoso, Chul-Moon Yoo and Taishi Ikeda for valuable comments and discussions. SJ thanks I. H. Dwivedi for discussions.
This work was partially supported by JSPS
KAKENHI Grant No. JP26400282 (T.H.).

\color{black}


\begin{thebibliography}{99}

\bibitem{GW150914} B. P. Abott et al. \emph{(LIGO Scientific Collaboration and Virgo Collaboration)}, Phys. Rev. Lett. {\bf 116}, 061102 (2016).

\bibitem{GW151226} B. P. Abott et al. \emph{(LIGO Scientific Collaboration and Virgo Collaboration)}, Phys. Rev. Lett. {\bf 116}, 241103 (2016).

\bibitem{GW170104} B. P. Abott et al. \emph{(LIGO Scientific Collaboration and Virgo Collaboration)}, Phys. Rev. Lett. {\bf 118}, 221101 (2017).

\bibitem{GW170814} B. P. Abott et al. \emph{(LIGO Scientific Collaboration and Virgo Collaboration)}, Phys. Rev. Lett. {\bf 119}, 141101 (2017).

\bibitem{GW170817} B. P. Abott et al. \emph{(LIGO Scientific Collaboration and Virgo Collaboration)}, Phys. Rev. Lett. {\bf 119}, 161101 (2017).

\bibitem{mmess} B. P. Abott et al. \emph{(LIGO Scientific Collaboration and Virgo Collaboration)}, Astrophys. J. Letters {\bf 848}, 1 (2017).

\bibitem{pen65} R. Penrose, Phys. Rev. Lett {\bf 14}, 57 (1965). 

\bibitem{Haw67} S. W. Hawking, Proc. R. Soc. Lond. {\bf A300}, 187 (1967).

\bibitem{reviews} R. Geroch and G. T. Horowitz, in \emph{General Relativity: Ans Einstein Centenary Survey}, Ed. S.W. Hawking and W. Israel, Cambridge University Press (1979); J. M. M. Sinovilla, Gen. Relav. Grav. {\bf 29}, 701 (1997). 

\bibitem{penrose1969} 
  R.~Penrose,
  Riv.\ Nuovo Cim.\  {\bf 1}, 252 (1969)
  [Gen.\ Rel.\ Grav.\  {\bf 34}, 1141 (2002)].

\bibitem{Oppenheimer-Snyder}
J.~R.~Oppenheimer and H.~Snyder,
  Phys.\ Rev.\  {\bf 56}, 455 (1939).

\bibitem{eardley-smarr1978} 
  D.~M.~Eardley and L.~Smarr,
  Phys.\ Rev.\ D {\bf 19}, 2239 (1979).
\bibitem{roberts1989} 
  M.~D.~Roberts,
  Gen.\ Rel.\ Grav.\  {\bf 21}, 907 (1989).
\bibitem{Ori-piran1987} 
  A.~Ori and T.~Piran,
  Phys.\ Rev.\ Lett.\  {\bf 59}, 2137 (1987).
\bibitem{Christodoulou1984} 
  D.~Christodoulou,
  Commun.\ Math.\ Phys.\  {\bf 93}, 171 (1984).
\bibitem{Joshi-Dwivedi1993} 
  P.~S.~Joshi and I.~H.~Dwivedi,
  Phys.\ Rev.\ D {\bf 47}, 5357 (1993).
  
\bibitem{Harada98} 
  T.~Harada,
  Phys.\ Rev.\ D {\bf 58}, 104015 (1998).

\bibitem{HaradaMaeda2001} T.~Harada and H.~Maeda,
  Phys.\ Rev.\ D {\bf 63}, 084022 (2001).
  
\bibitem{coop}
 F.~I.~Cooperstock, S.~Jhingan, P.~S.~Joshi and T.~P.~Singh,
  Class.\ Quant.\ Grav.\  {\bf 14}, 2195 (1997)

\bibitem{Qs-CS} S. S. Deshingkar, S. Jhingan, P. S. Joshi, Gen. Rel. Grav. {\bf 30} 1477 (1998); S. M. C. V. Goncalves, S. Jhingan Int. J. Mod. Phys. {\bf D11}, 1469 (2002).
 
\bibitem{ChakrabartiJoshi1994} 
  S.~K.~Chakrabarti and P.~S.~Joshi,
  Int.\ J.\ Mod.\ Phys.\ D {\bf 3}, 647 (1994).

\bibitem{Singh1998} 
  T.~P.~Singh,
  Gen.\ Rel.\ Grav.\  {\bf 30}, 1563 (1998).

\bibitem{HiscockWilliamsEardley1982} 
  W.~A.~Hiscock, L.~G.~Williams and D.~M.~Eardley,
  Phys.\ Rev.\ D {\bf 26}, 751 (1982).

\bibitem{VazWitten1993} 
  C.~Vaz and L.~Witten,
  Phys.\ Lett.\ B {\bf 325}, 27 (1994).

\bibitem{VazWitten1995} 
  C.~Vaz and L.~Witten,
  Class.\ Quant.\ Grav.\  {\bf 12}, 2607 (1995).
  
\bibitem{VazWitten1996} 
  C.~Vaz and L.~Witten,
  Class.\ Quant.\ Grav.\  {\bf 13}, L59 (1996)

\bibitem{Barve+1998-1} 
  S.~Barve, T.~P.~Singh, C.~Vaz and L.~Witten,
  Nucl.\ Phys.\ B {\bf 532}, 361 (1998).

\bibitem{Barve+1998-2}
  S.~Barve, T.~P.~Singh, C.~Vaz and L.~Witten,
  Phys.\ Rev.\ D {\bf 58}, 104018 (1998).

\bibitem{VazWitten1998} 
  C.~Vaz and L.~Witten,
  Phys.\ Lett.\ B {\bf 442}, 90 (1998).

\bibitem{HaradaIguchiNakao2000-1} 
  T.~Harada, H.~Iguchi and K.~i.~Nakao,
  Phys.\ Rev.\ D {\bf 61}, 101502 (2000).

\bibitem{HaradaIguchiNakao2000-2} 
  T.~Harada, H.~Iguchi and K.~i.~Nakao,
  Phys.\ Rev.\ D {\bf 62}, 084037 (2000).

\bibitem{Harada+2001} 
  T.~Harada, H.~Iguchi, K.~i.~Nakao, T.~P.~Singh, T.~Tanaka and C.~Vaz,
  Phys.\ Rev.\ D {\bf 64}, 041501 (2001).

\bibitem{IguchiHarada2001} 
  H.~Iguchi and T.~Harada,
  Class.\ Quant.\ Grav.\  {\bf 18}, 3681 (2001).
  
\bibitem{HaradaIguchiNakao2002} 
  T.~Harada, H.~Iguchi and K.~i.~Nakao,
  Prog.\ Theor.\ Phys.\  {\bf 107}, 449 (2002).

\bibitem{IguchiNakaoHarada1998} 
  H.~Iguchi, K.~i.~Nakao and T.~Harada,
  Phys.\ Rev.\ D {\bf 57}, 7262 (1998).

\bibitem{IguchiHaradaNakao1999} 
  H.~Iguchi, T.~Harada and K.~i.~Nakao,
  Prog.\ Theor.\ Phys.\  {\bf 101}, 1235 (1999).

\bibitem{IguchiHaradaNakao2000} 
  H.~Iguchi, T.~Harada and K.~i.~Nakao,
  Prog.\ Theor.\ Phys.\  {\bf 103}, 53 (2000).

\bibitem{Barrabes-Hogan-book}
 C.~Barrab\'es and P.~A.~Hogan, \textit{Singular Null Hypersurface in General Relativity}, World Scientific (2003), Chapter 5.
  
\bibitem{toolkit}
E.~Poisson, \textit{A Relativist's Toolkit}, Cambridge University Press (2007), Chapter 3,\\
E. Poisson, arXiv:gr-qc/0207101.

\bibitem{carr05}
  B. J. Carr and A. A. Coley, Gen. Rel. Grav. {\bf 37} 2165 (2005).

\bibitem{Barve+1999} 
  S.~Barve, T.~P.~Singh, C.~Vaz and L.~Witten,
  Class.\ Quant.\ Grav.\  {\bf 16}, 1727 (1999).
\bibitem{Duffy-Nolan2011} 
  E.~M.~Duffy and B.~C.~Nolan,
  arXiv:1108.1103 [gr-qc].

\bibitem{Jhingan-Dwivedi-Barve2010} 
  S.~Jhingan, I.~H.~Dwivedi and S.~Barve,
  Phys.\ Rev.\ D {\bf 84}, 024001 (2011). 

 \bibitem{positivity-proof} 
From \eq{orange-explicit} and \eq{orange-limit}, we have 
\begin{align}
z=\frac{1-\frac{az}{3}}{(1-az)^{1/3}}. \label{z-const}
\end{align}
$(1-az)^{-1/3}\geq 1$ holds because of $0\leq az<1$. For $X\geq 1$, one gets the inequality $X^{-1/3}\geq X^{-1}$.
Thus, we get the inequality for $z$: 
\begin{align}
z=\frac{1-\frac{az}{3}}{(1-az)^{1/3}}\geq \frac{1-\frac{az}{3}}{1-az} \geq1. \label{z-greater-1}
\end{align}
On the other hand, if the coefficient of the second term in \eq{E-shell-dust-sch} can be negative, then by recasting the coefficient with using the relation \eq{z-const}, the condition for the negativity can be simply written as $z^3<1-a^2z^2/9$, {\it i.e.}, $z<1$. However, according to \eq{z-greater-1}, we have $z\geq 1$. This proves the non-negativity for the coefficient of the second term in \eq{E-shell-dust-sch}.
\end{thebibliography}
\end{document}